\documentclass[12pt]{article}

\usepackage{graphicx}
\graphicspath{{./figures/}}
\usepackage{dcolumn}
\usepackage{bm}
\usepackage{amsfonts}
\usepackage{amsthm}
\usepackage{amsmath}
\usepackage{amssymb}
\usepackage{slashed}
\usepackage{color}
\usepackage{textcomp}
\usepackage{subfig}
\usepackage{cite}

\def\be{\begin{equation}}
\def\ee{\end{equation}}
\def\ba{\begin{eqnarray}}
\def\ea{\end{eqnarray}}

\def\bdm{\begin{displaymath}}
\def\edm{\end{displaymath}}
\def\la{~\mbox{\raisebox{-.6ex}{$\stackrel{<}{\sim}$}}~}

\def\bq{\begin{quote}}
\def\eq{\end{quote}}

 at 10truept

\newcommand{\Mpl}{M_{\rm P}}

\newcommand{\bea}{\begin{eqnarray}}
\newcommand{\eea}{\end{eqnarray}}

\newcommand{\bi}{\begin{itemize}}
\newcommand{\ei}{\end{itemize}}

\newcommand{\beq}{\begin{equation}}
\newcommand{\eeq}{\end{equation}}
\newcommand{\beqa}{\begin{eqnarray}}
\newcommand{\eeqa}{\end{eqnarray}}

\def\la{~\mbox{\raisebox{-.6ex}{$\stackrel{<}{\sim}$}}~}

\def\ltap{\ \raise.3ex\hbox{$<$\kern-.75em\lower1ex\hbox{$\sim$}}\ }
\def\gtap{\ \raise.3ex\hbox{$>$\kern-.75em\lower1ex\hbox{$\sim$}}\ }
\def\gl{\ \raise.5ex\hbox{$>$}\kern-.8em\lower.5ex\hbox{$<$}\ }
\def\roughly#1{\raise.3ex\hbox{$#1$\kern-.75em\lower1ex\hbox{$\sim$}}}

\begin{document}

\thispagestyle{empty}
\begin{flushright}
DESY 19-130\\
July 12, 2019
\end{flushright}
\vspace*{1.5cm}

\begin{center}
{\Large \bf A Goldilocks Higgs}

\vskip.3cm

\vspace*{1.26cm} {\large Nemanja Kaloper$^{a,}$\footnote{\tt
kaloper@physics.ucdavis.edu} and  Alexander Westphal$^{b,}$\footnote{\tt alexander.westphal@desy.de}
}\\
\vspace{.5cm} $^a${\em Department of Physics, University of
California, Davis, CA 95616, USA}\\
\vspace{.5cm} $^b${\em Deutsches Elektronen-Synchrotron DESY, Theory Group, D-22603 Hamburg, Germany}\\

\vspace*{1.26cm} 
ABSTRACT
\end{center}

The Higgs could couple to a topological 4-form sector which yields a complex vacuum structure. In general such couplings could lead to direct CP violation in the Higgs sector. In many of the Higgs vacua electroweak symmetry is unbroken. In just as many it breaks when the
4-form flux is large enough. For a fixed value of flux, the symmetry breaking vacua have a smaller vacuum energy than the symmetric
ones, where the difference is quantized because it is set by the $4$-form flux. This leads to the possibility 
that there is a value of the 4-form flux for any UV contributions to the Higgs 
{\it vev} that automatically cancels it down to the right value, $\sim$ TeV, if the 4-form charges are quantized in the units of the  electroweak scale. This would still leave the cosmological constant which could be selected anthropically.

\vfill \setcounter{page}{0} \setcounter{footnote}{0}
\newpage

\section{Introduction}

Why is the Standard Model (SM) so light? In the standard formulation of the SM, the masses of all particles, including the Higgs, are set by the Higgs {\it vev} after the electroweak (EW) spontaneous symmetry breaking (SSB). However, if there are any additional 
heavy degrees of freedom in the universe, which could either play a role in the unification of forces or be dark matter, that have non-ignorable couplings to the Higgs, the Higgs {\it vev} would receive significant contributions from their vacuum fluctuations. If string theory is right, this seems inevitable, since all of  light QFT should be understood as an EFT with many heavy states integrated out. Without an obvious cancellation mechanism as for example low scale SUSY, it is puzzling to see that such heavy states do not influence the observed low energy physics more significantly.

The SM, therefore, appears to be special. Its specialty is tracked down to the special value of a single dimensional number, the Higgs {\it vev}. This evokes an obvious analogy to another very special dimensional parameter characterizing the universe, the cosmological constant. A possible way to understand its smallness is to imagine that it is a characteristic of a nontrivial structure of the manifold of vacua in the theory, rather than an unpredictable parameter in a simple vacuum. The Higgs {\it vev} might be similar. We might not have a single vacuum hosting 
the light low energy theory, enforced by local dynamics controlled by symmetries which preclude UV contaminations. Instead we may have a large multiplicity of Higgs vacua where the possible {\it vevs} are quantized with a step size $\sim$ TeV. In this case, the observed features of the SM could be understood as a cosmological coincidence without any need for new dynamics and new particles.  The Higgs {\it vev} selection would be very weakly anthropic, while the cosmological constant in these vacua could then be selected by the more refined cosmological evolution combined with anthropic boundary conditions.

In this communication we will present a framework which realizes such a scenario. The idea is simple: imagine that
the Higgs {in general might not be } a CP-even state, and that it has an `axion'-like admixture, {and let it couple} to a 4-form sector.
If this is the case, the Higgs effective potential could include terms such as
\be
V \ni \frac{c}{24} \epsilon_{\mu\nu\lambda\sigma} F^{\mu\nu\lambda\sigma} |\phi|^2
\label{mixing}
\ee
such as those which appear in 4D flux monodromy models of inflation \cite{Kaloper:2008fb,Kaloper:2008qs,Kaloper:2011jz,Kaloper:2016fbr,DAmico:2017cda}, but where $c$ could be a number of indefinite CP\footnote{The specific CP properties of $c$ depend on the UV completion.  For example if we start with a compactification which includes terms $\propto G_{\mu\nu\lambda\sigma} F^{\mu\nu\lambda\sigma} |\phi|^2$, where $G$ and $F$ are 
two different $4$-forms, and $G$ has a magnetic flux whose discharge is much more suppressed than the $F$ flux,
there resulting $4D$ term will be (\ref{mixing}) where $c$ is CP-odd. However there could be 
configurations with terms $\propto G^2 \epsilon_{\mu\nu\lambda\sigma}  F^{\mu\nu\lambda\sigma}  |\phi|^2$ which break CP, and thus a general construction may involve both types of contributions.}. Since the whole action should
be invariant, and the term $\epsilon F$ is CP-odd, this means that in principle $|\phi|^2$ is also CP-indefinite. 
E.g., if we fix $c$ to be CP-even, then the field which couples to $\epsilon F$ would have to be an axion-like mode,
if $c$ were CP-odd, the field coupling to $\epsilon F$ would be a CP-even scalar and so on. Here we instead assume the general case, and consider a $c$ which is {\it a priori} CP-indefinite. Note that terms like this have been used 
previously to set up landscapes of of $\theta_{QCD}$ and gauge hierarchy \cite{nicolai,Dvali:2001sm,Dvali:2003br,Dvali:2004tma,Dvali:2005an} and have been {found} to give new angles of attack on the cosmological constant problem \cite{nicolai,Dvali:2001sm,Brown:1987dd,Brown:1988kg,Bousso:2000xa,Polchinski:2006gy}. Specifically, Higgs-$4$-form couplings were investigated in \cite{Dvali:2003br,Dvali:2004tma}.

Since the 4-form field strength
$F_{\mu\nu\lambda\sigma} = 4 \partial_{[\mu} A_{\nu\lambda\sigma]}$ is dimension-2, this term is actually  renormalizable, and aside from {allowing}  the Higgs to be of indefinite CP, it could have been included in the SM from the start. However, since the 4-form field strength is additive, $F \sim N q$ where $N$ is an integer and $q$ the charge of a membrane sourcing the flux of $F$, the field strengths could easily be very large, of either sign. 
Further, it
will take discrete values, separated by a unit of $q$, leading to many different low energy theories of the
Higgs sector. 

Thus any UV contributions to the Higgs {\it vev} can be compensated by the 4-form flux in some of the vacua, however large these corrections might be. If the unit of charge is picked to be $c q \sim {\rm TeV}^2$, the flux of the 4-form will be just right to cancel the large UV corrections for some value of $N$.

\section{A Christmas Tree of Electroweak Vacua}

Let us now flesh out the details. For simplicity we only work with the Higgs sector. The idea is to assume that the 
Higgs potential includes the standard terms, 
\be
V_0 = \frac{\lambda}{4} |\phi|^4 - \frac{v^2}{2}  |\phi|^2 + \Lambda
\ee
where however $v$ can be arbitrarily large, including
any possible UV contribution from above the EW scale; $\Lambda$ is an a priori arbitrary contribution to the
cosmological constant  in the vacuum. We will take it to be a globally dynamical variable, similar to \cite{Bousso:2000xa,Polchinski:2006gy}.  
In addition, we add the terms 
`monodromizing' the Higgs {\it vev}, 
\be
\Delta V = \frac{c}{24}  \epsilon_{\mu\nu\lambda\sigma} F^{\mu\nu\lambda\sigma} |\phi|^2 - \frac{1}{48} F_{\mu\nu\lambda\sigma}^2 
\ee
and then we dualize $F$, replacing it with its dual magnetic field strength. This amounts to adding
\be
\delta V =   \frac{Q}{24}  \epsilon_{\mu\nu\lambda\sigma}  (F^{\mu\nu\lambda\sigma} 
-  4 \partial^{[\mu} A^{\nu\lambda\sigma]})
\ee
to the effective potential. Adding them up gives our modification to the SM Lagrangian,
\be
\delta {\cal L} =  - \frac{1}{48} F_{\mu\nu\lambda\sigma}^2 + \frac{Q  + c |\phi|^2}{24}  \epsilon_{\mu\nu\lambda\sigma}  F^{\mu\nu\lambda\sigma} 
+  \frac16  \epsilon_{\mu\nu\lambda\sigma}  (\partial^{\mu} Q) A^{\nu\lambda\sigma}  \quad . 
\ee
The last term says that locally $Q$ is a constant, which can only change by a membrane emission, 
since a membrane with a charge $q$ couples to $A$ by 
\be
\frac{q}{6} \int d^3 \xi \sqrt{\gamma} e^{abc} \partial_a x^\mu  \partial_b x^\nu  \partial_c x^\lambda A_{\mu\nu\lambda} \quad . 
\ee
This means that $Q$ is quantized in the units of
the membrane charge $q$, $Q = Nq$. Now completing the square in $F$ and integrating it out yields the final formula for the extended Higgs potential,
\be
V = \frac{\lambda}{4} |\phi|^4 - \frac{v^2}{2} |\phi|^2 + \frac12 (Q + c |\phi|^2)^2 + \Lambda = \frac{\bar \lambda}{4} |\phi|^4 - \frac{\bar v^2}{2} |\phi|^2 + \frac12 Q^2 + \Lambda
\ee
where
\ba
\bar \lambda &=& \lambda + 2c  \\
\bar v^2 &=& v^2 - 2c Q  \quad .
\ea
Note that the effective potential (7) resembles the relaxion \cite{gkr,choi} realized via a monodromy \cite{ibanez}, 
which was `frozen' out and replaced by a locally constant value $Q$. Such a model was explored in \cite{Herraez:2016dxn}. 

The formula in Eq. (9) is particularly important. Since $Q = Nq$, no matter what $v^2$ is, we can always pick an integer
$N$ such that $\bar v^2$ is in the TeV window required to keep the SM at the observed scales. This means
that in order to make this natural, and avoid gross fine tunings we must pick 
\be
cq \la {\rm TeV}^2 \quad . 
\ee
In this case, we are guaranteed that there is a vacuum branch for any $v \gg$ TeV, such that
\be
N_* = \Bigl[\frac{v^2}{cq} \Big]
\ee 
is the integer closest to the ratio $v^2/cq$ from below, for any
$v^2$. In other words, whatever the UV physics that could affect the Higgs {\it vev}, there is a flux of $F$ that 
compensates it, retaining the expectation value of the Higgs in the EW window. 

Note, that in this case the neighboring values of the flux, $N= N_* \pm 1$ are already problematic from the low energy point of view.
For $N = N_*+1$, the EW symmetry is restored, since $v^2 - 2cq (N_*+1)$ flips sign, as the flux overcompensates
$v^2$. For $N = N_* -1$, the Higgs {\it vev} is larger by ${\cal O}(1)$, rendering the SM particles all heavier, while
all the charges remain fixed. This is problematic for low energy physics, in particular BBN. If we require that the universe should evolve to allow nontrivial very low energy dynamics instead of being a cold boring place, this clearly favors\footnote{We are assuming that the Yukawa couplings to the EW fermions are fixed to their observed values reflecting the observed reality. This is our {\it prior}, which for example excludes the limit of the `weakless' universe \cite{weakless}.}  the critical flux with $N_*$. 

Further issues -- and insights -- arise when we consider the cosmological constant contribution from EW SSB. 
In  vacua with $N > N_*$  for a fixed $v^2$, we have {$\bar v^2 < 0$} and there is no EW SSB. The SM in these states is completely massless, relativistic, yet with the net vacuum energy given by
\be
\frac12 N^2 q^2 + \Lambda \quad.
\ee
Such universes are very inhospitable. They can only have radiation being inflated away forever, or rapidly crunching up if $\Lambda$ is sufficiently negative.

In contrast, in the vacua with $N < N_*$, EW SSB takes place in the IR, when the universe cools down, 
and the effective cosmological constant in these states is 
\be
\Lambda_{N} = \frac12 N^2 q^2 + \Lambda - \frac14 \frac{( v^2 - 2 N cq)^2}{\bar \lambda}\quad. 
\ee
If we compare the cosmological constants in the preferred state $N_*$ and the state $N_*-1$ right next to it, we
find that
\be\label{eq:CCspacing}
\Delta \Lambda = \Lambda_{N_*} - \Lambda_{N_*-1} \simeq \frac{q v^2}{c} (1- \frac{c}{\bar \lambda})  \quad  .
\ee 
Since $\bar \lambda = \lambda + 2c$ -- and $c$ includes the CP-violating effects in the Higgs sector -- it is natural to expect that
\be
c_{\tt CP-even}/\bar \lambda \ll 1 \quad .
\ee
This implies that the cosmological constant in the state with $N_*-1$, adjacent to the conventional SM, is much smaller. The cosmological dynamics which picks the late low energy state of the universe, introducing dynamics in the additional cosmological constant term $\Lambda$, as in for example \cite{Bousso:2000xa,Polchinski:2006gy,weinberg,Wein}, would therefore have to pick the state with $N_*$, since the adjacent states either

\begin{itemize}
\item  a) don't break EW symmetry or 
\item b) break it too badly, making SM too heavy.
\end{itemize}

\begin{figure*}
\centering
\includegraphics[scale=0.5]{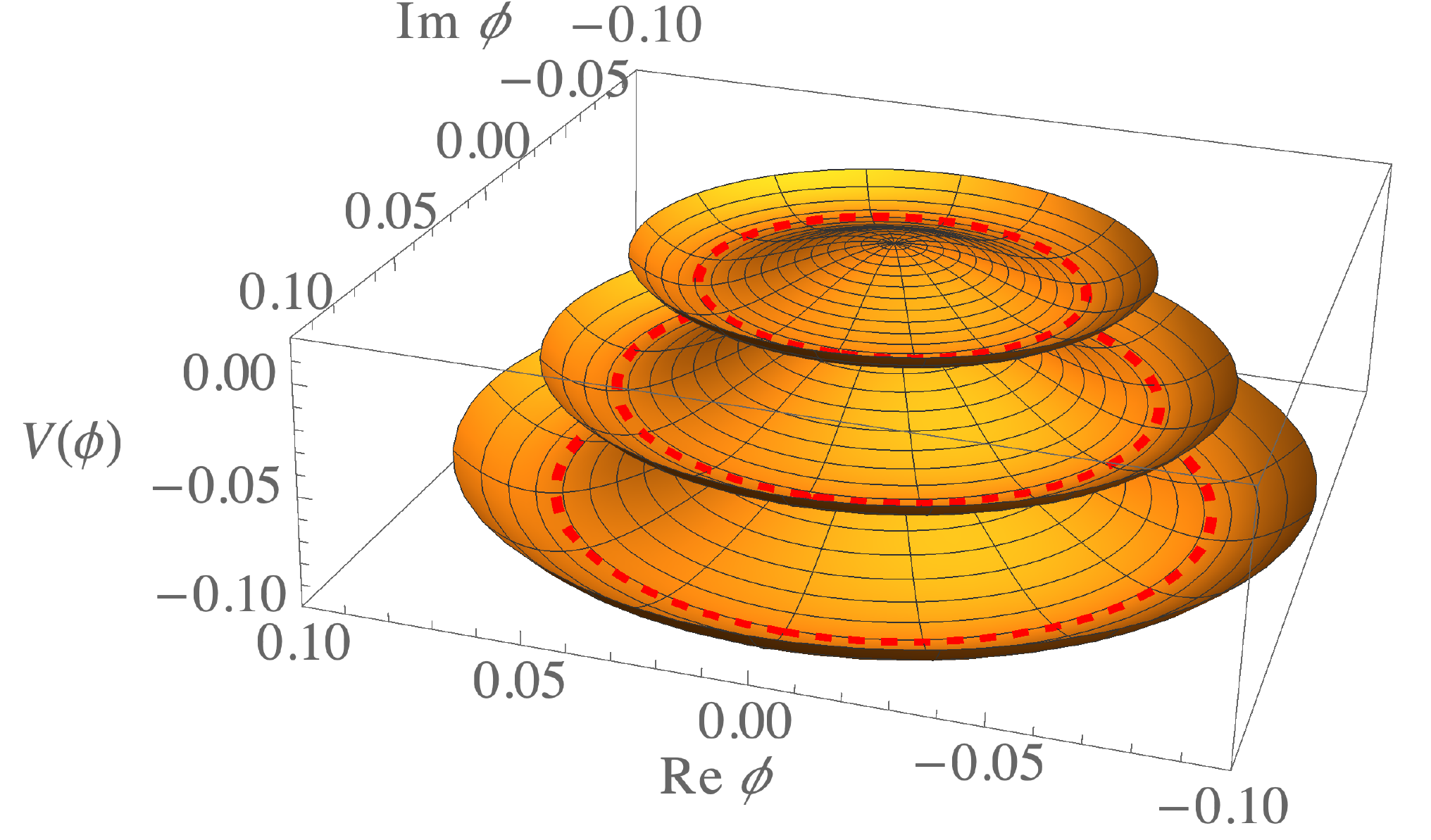}
\caption{The Christmas tree of EW SSB vacua for parameter choices $\lambda=1$, $c=0.1$, $q=0.01$. The red circles delineate the vacuum manifolds. The highest `goldilocks' branch with the smallest scale EW SSB vacuum has small positive cosmological constant, the branches below are deep AdS.}
\label{fig:ChristmasTree}
\end{figure*}

\noindent Further, since the adjacent state with $N_*-1$ has a much smaller cosmological constant, once {$\Lambda_{N_*}$} is selected to be 
\be
{\Lambda_{N_*} }\sim 10^{-122} \Mpl^4
\ee 
the cosmological constant in the state $N_*-1$ will be huge and negative.  The situation is depicted in Fig.~\ref{fig:ChristmasTree} for a choice of toy model parameters.

We note that the spacing of the cosmological constant values around zero $\Delta\Lambda\sim qv^2/c = \frac{1}{c^2} cq\,v^2 \sim {\rm TeV}^2 v^2$ is of the same scale as the scale of the residual cosmological constant in low-energy supersymmetry with $\Lambda_{MSSM} \sim m_{3/2}^2 \Mpl^2$ since there $v\sim m_{3/2}\sim {\rm TeV}$ and $M_{\rm GUT}<v<\Mpl$.  Hence, in our setup the scale of residual cosmological constant problem is reduced about as much as in models with low-energy supersymmetry as long as $c$ is not too small.

\begin{figure*}[t!]
\centering
\includegraphics[scale=0.38]{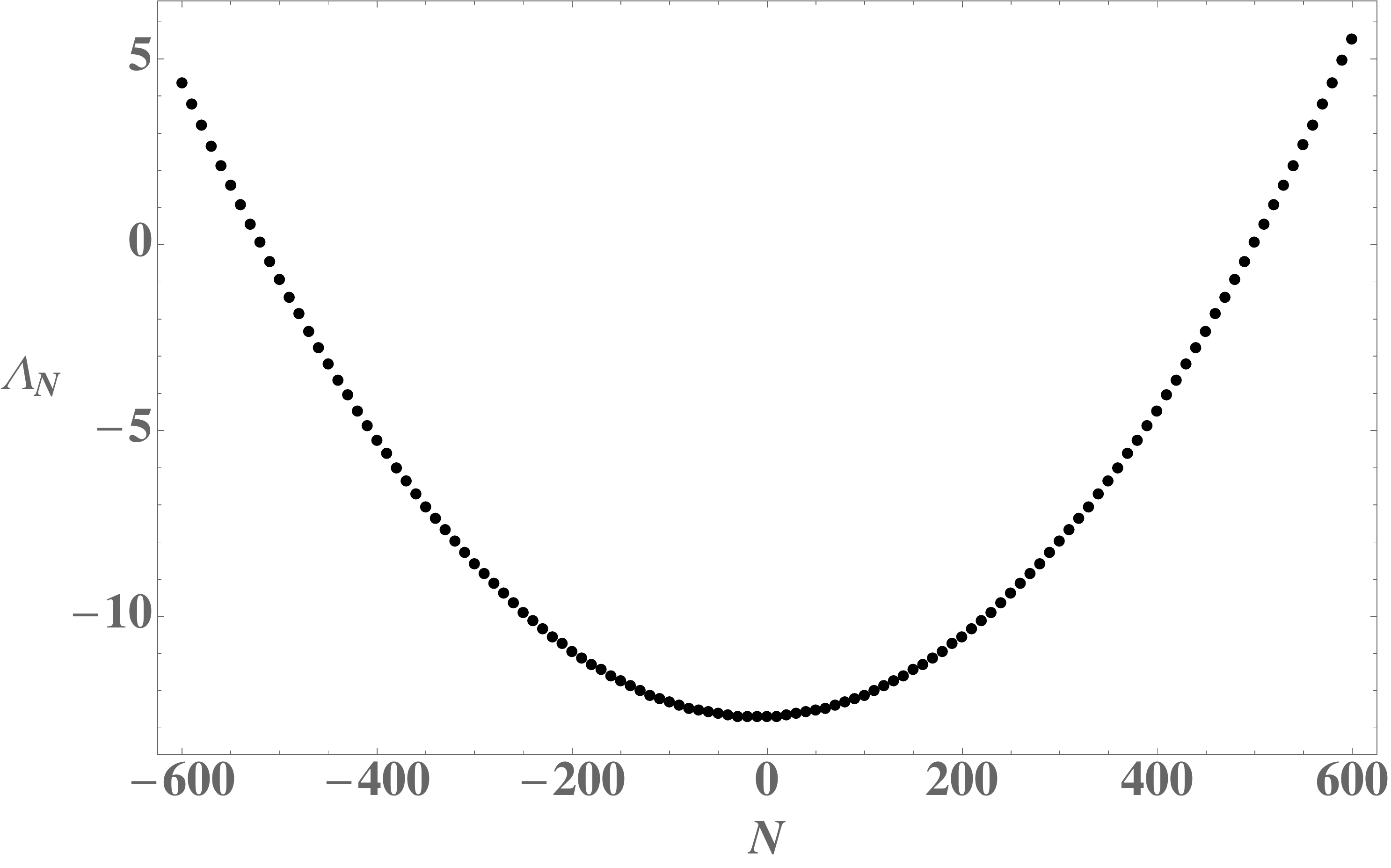}
\caption{The distribution of the cosmological constant in the SSB vacua for parameter choices $\lambda=1$, $c=0.1$, $q=0.01$. The values of the vacuum energy are shown for flux numbers $N$ in steps of 10.}
\label{fig:CCs}
\end{figure*}

This means, if our universe ever transitions to such a state
it will crunch immediately due to a huge negative vacuum energy. However, since a probability of such a transition is suppressed by
\be
P \sim \exp( -\frac{27\pi^2}{2}  \frac{\sigma^4}{(\Delta \Lambda)^3}) \sim  \exp( -\frac{27\pi^2}{2}  \frac{{c^3} \sigma^4}{q^3 v^6}) 
\ee
such disastrous transitions are extremely unlikely if the scale of the brane tension is controlled by the UV (that is, by scales of order $v$ so $\sigma \sim v^3$). In that case, 
\be
\sigma^4/v^6 \sim v^6 \gg q^3
\ee
and the state with $N_*$ units of flux is as stable as can be.
In the early universe, of course, transitions can and will occur more rapidly -- at larger values of $\Lambda$ -- which will populate states with $N_*$ units of flux early on. These will be the states which will support interesting low energy cosmology, explaining why we observe it \cite{weinberg,Wein}.  Note also from Eq. (9) that while initially the cosmological constant for the states $N < N_*$ decreases, becoming negative if $\Lambda_{N_*}$ is anthropically selected to be in the observed window, this quickly turns around and $\Lambda_N$ starts to grow again thanks to
$c< \bar \lambda$. We show the distribution of the cosmological constant for a numerical example in Fig.~\ref{fig:CCs}.

This means that the universe with the correct small Higgs {\it vev} is really a rather special place.
Most of the other universes in our framework have a much larger cosmological constant, with either too large a Higgs {\it vev} or without  EW SSB. There may be some universes ``nearby'' where the Higgs {\it vev} is ${\cal O}(1)$ larger, that would be a problem for cosmochemistry. However those few universes would have a large negative cosmological constant once the SM one is selected, and hence are doubly disfavored. 

\section{Discussion}

In summary, we have found that a nontrivial coupling of the Higgs to topological sectors -- modeled here
by a 4-form fluxes and a monodromy-like mixing of the Higgs {\it vev} with it -- can generate a mini-multiverse
of Higgs vacua. Many of them yield wrong low energy dynamics for the SM. The SM is either too heavy, or EW SSB never happens. However, if the charge of the membrane sourcing the 4-form flux is set by the EW scale, there is always one vacuum, supporting the right SM which can get a small cosmological constant by cosmological evolution obeying anthropic boundary conditions. The couplings to the 4-form which allow for this are in fact power counting renormalizable, preserving the SM dynamics to the tee. {However, new physics can enter if we assume that the Higgs is a CP-indefinite state, so that we allow for a general coupling which} breaks CP explicitly. This might be a way to experimentally probe the proposal -- or at least, to constrain it, by finding that the Higgs sector does not break CP. Another possible test is cosmology. The relaxation dynamics of the cosmological constant requires nucleation  of 
bubbles, and if the charges are set by the EW scale, the creation of such bubbles in the late universe might have affected the cosmological gravitational wave background. 

\vskip.5cm

{\bf Note Added}: While this manuscript was being written, ref.~\cite{Giudice:2019iwl} appeared which has significant overlap with our work.

\vskip.5cm

{\bf Acknowledgments}: 
We would like to thank G. D'Amico and A. Lawrence for many 
very useful discussions. We also thank Brothers Grimm for inspiration. NK thanks CERN Theory Division, DESY Theory Group, Mainz MITP and KITP, UCSB for kind hospitality in the course of this work. AW thanks CERN Theory Division for kind hospitality in the course of this work. NK is supported in part by the DOE Grant DE-SC0009999. AW is supported by the ERC Consolidator Grant STRINGFLATION under the HORIZON 2020 grant agreement no. 647995, as well as by the Deutsche Forschungsgemeinschaft (DFG, German Research Foundation) under Germany's Excellence Strategy -- EXC 2121 ``Quantum Universe'' -- 390833306.


\begin{thebibliography}{99}


\bibitem{Kaloper:2008fb}
  N.~Kaloper and L.~Sorbo,
  ``A Natural Framework for Chaotic Inflation,''
  Phys.\ Rev.\ Lett.\  {\bf 102}, 121301 (2009).
\bibitem{Kaloper:2008qs} 
  N.~Kaloper and L.~Sorbo,
  ``Where in the String Landscape is Quintessence,''
  Phys.\ Rev.\ D {\bf 79}, 043528 (2009).
\bibitem{Kaloper:2011jz}
  N.~Kaloper, A.~Lawrence and L.~Sorbo,
  ``An Ignoble Approach to Large Field Inflation,''
  JCAP {\bf 1103}, 023 (2011).
\bibitem{Kaloper:2016fbr} 
  N.~Kaloper and A.~Lawrence,
  ``London equation for monodromy inflation,''
  Phys.\ Rev.\ D {\bf 95}, no. 6, 063526 (2017).
\bibitem{DAmico:2017cda} 
  G.~D'Amico, N.~Kaloper and A.~Lawrence,
  ``Monodromy Inflation in the Strong Coupling Regime of the Effective Field Theory,''
  Phys.\ Rev.\ Lett.\  {\bf 121}, no. 9, 091301 (2018). 


\bibitem{nicolai} 
  A.~Aurilia, H.~Nicolai and P.~K.~Townsend,
  ``Hidden Constants: The Theta Parameter of QCD and the Cosmological Constant of N=8 Supergravity,''
  Nucl.\ Phys.\ B {\bf 176}, 509 (1980).
      


\bibitem{Dvali:2001sm} 
  G.~R.~Dvali and A.~Vilenkin,
  ``Field theory models for variable cosmological constant,''
  Phys.\ Rev.\ D {\bf 64}, 063509 (2001).
\bibitem{Dvali:2003br} 
  G.~Dvali and A.~Vilenkin,
  ``Cosmic attractors and gauge hierarchy,''
  Phys.\ Rev.\ D {\bf 70}, 063501 (2004).
\bibitem{Dvali:2004tma} 
  G.~Dvali,
  ``Large hierarchies from attractor vacua,''
  Phys.\ Rev.\ D {\bf 74}, 025018 (2006).
\bibitem{Dvali:2005an} 
  G.~Dvali,
  ``Three-form gauging of axion symmetries and gravity,''
  hep-th/0507215.

\bibitem{Dvali:2005zk} 
  G.~Dvali,
  ``A Vacuum accumulation solution to the strong CP problem,''
  Phys.\ Rev.\ D {\bf 74}, 025019 (2006).


\bibitem{Brown:1987dd} 
  J.~D.~Brown and C.~Teitelboim,
  ``Dynamical Neutralization of the Cosmological Constant,''
  Phys.\ Lett.\ B {\bf 195}, 177 (1987).
\bibitem{Brown:1988kg} 
  J.~D.~Brown and C.~Teitelboim,
  ``Neutralization of the Cosmological Constant by Membrane Creation,''
  Nucl.\ Phys.\ B {\bf 297}, 787 (1988).

        
\bibitem{Bousso:2000xa} 
  R.~Bousso and J.~Polchinski,
  ``Quantization of four form fluxes and dynamical neutralization of the cosmological constant,''
  JHEP {\bf 0006}, 006 (2000).
\bibitem{Polchinski:2006gy} 
  J.~Polchinski,
  ``The Cosmological Constant and the String Landscape,''
  hep-th/0603249.

 
\bibitem{gkr} 
  P.~W.~Graham, D.~E.~Kaplan and S.~Rajendran,
  ``Cosmological Relaxation of the Electroweak Scale,''
  Phys.\ Rev.\ Lett.\  {\bf 115}, no. 22, 221801 (2015).
  
\bibitem{choi} 
  K.~Choi and S.~H.~Im,
  ``Realizing the relaxion from multiple axions and its UV completion with high scale supersymmetry,''
  JHEP {\bf 1601}, 149 (2016).
 
\bibitem{ibanez} 
  L.~E.~Ibanez, M.~Montero, A.~Uranga and I.~Valenzuela,
  ``Relaxion Monodromy and the Weak Gravity Conjecture,''
  JHEP {\bf 1604}, 020 (2016).
  
\bibitem{Herraez:2016dxn} 
  A.~Herraez and L.~E.~Ibanez, ``An Axion-induced SM/MSSM Higgs Landscape and the Weak Gravity Conjecture,''
  JHEP {\bf 1702}, 109 (2017).

\bibitem{weakless}
  R.~Harnik, G.~D.~Kribs and G.~Perez,
  ``A Universe without weak interactions,''
  Phys.\ Rev.\ D {\bf 74}, 035006 (2006). 
  
  
\bibitem{weinberg}
S.~Weinberg,
``The Cosmological Constant Problem,''
Rev.\ Mod.\ Phys.\  {\bf 61}, 1 (1989).

       
  
    
\bibitem{Wein} 
  S.~Weinberg,
  ``Anthropic Bound on the Cosmological Constant,''
  Phys.\ Rev.\ Lett.\  {\bf 59}, 2607 (1987).

   
  
   
\bibitem{Giudice:2019iwl} 
  G.~F.~Giudice, A.~Kehagias and A.~Riotto,
  ``The Selfish Higgs,''
  arXiv:1907.05370 [hep-ph].
  
             
\end{thebibliography}
\end{document}